Domain-resolved room-temperature magneto-electric coupling in manganite-titanate heterostructures

R.V. Chopdekar[1], M.Buzzi[2], C.A. Jenkins[3], E. Arenholz[3], F. Nolting[2], and Y. Takamura[1]

1 Dept. of Chemical Engineering & Materials Science, Univ. of California, Davis

2 Swiss Light Source, Paul Scherrer Institute

3 Advanced Light Source, Lawrence Berkeley National Laboratory

We present a model artificial multiferroic system consisting of a (011)-oriented ferroelectric Pb(Mg,Nb,Ti)$O_3$ substrate intimately coupled to a ferromagnetic (La,Sr)Mn$O_3$ film through epitaxial strain and converse piezoelectric effects. Electric field pulse sequences of less than 6 kV/cm were shown to induce large reversible and bistable remanent strains in the manganite film. Magnetic hysteresis loops demonstrate that the changes in strain states result in significant changes in magnetic anisotropy from a highly anisotropic two-fold magnetic symmetry to a more isotropic one. Such changes in magnetic anisotropy are reversible upon multiple cycles and are stable at zero applied electric field, and are accompanied by large changes in resistivity. We directly image the change between the two-fold and isotropic magnetic configurations at the scale of a single ferromagnetic domain using X-ray photoemission electron microscopy as a function of applied electric field pulses. Imaging the domain reversal process as a function of electric field shows that the energy barrier for magnetization reversal is drastically lowered, by up to 70% as determined from free energy calculations, through the anisotropic strain change generated by the ferroelectric substrate. Thus, an electric field pulse can be used to 'set' and 'reset' the magnetic anisotropy orientation and resistive state in the film, as well as lowering the coercive field required to reverse magnetization, showing a promising route towards electric-field manipulation of multifunctional nanostructures at room temperature.

While single phase multiferroic materials exist in nature, often they have relatively weak magnetoelectric coupling and their ordering temperatures are well below ambient conditions. In an effort to develop room-temperature magnetoelectric functionality, there has been a strong thrust of research in the direction of artificial multiferroics consisting of composite structures.[1,2] A magnetoelectric property from cross coupling of different ferroic phases may arise due to a structural interaction as in magnetostrictive-piezoelectric composites,[3] or may be produced from an electric charge-based phenomenon such as through the use of a gate voltage in proximity to a ferromagnetic channel.[4] This emergent magnetoelectric functionality has been recently implemented in so-called 'straintronic' device designs to minimize active power consumption as compared to more conventional current-assisted nanomagnet memory writing schemes.[5,6]

Epitaxial ferromagnetic manganite films have often been used as prototypical materials showing dynamic tuning of functional properties such as magnetization or resistivity when coupled to ferroelectric substrates such as $BaTiO_3$.[7-9] Spatially resolved studies of the manganite-$BaTiO_3$ system illustrate that by changing strain magnitude and symmetry through traversing $BaTiO_3$ phase transitions,[10] a persistent 45° or 90° rotation of magnetic anisotropy occurs at the scale of individual magnetic domains and suggests that similar artificial multiferroic systems employing manganite layers are ideal candidates to explore electric field-induced magnetization rotation or reversal of nanoscale magnetic elements. Additionally, devices utilizing doped manganites have shown current-driven switching of magnetization as well as nearly full spin polarization.[11,12] Thus, the strong link between structure, ferromagnetism, and spin polarization in manganites can be used to influence not only magnetization orientation, but can also be an avenue for voltage control of spin transport in heterostructures.

Recent work has shown that a large, anisotropic, and hysteretic in-plane strain may be generated from the (011) crystal orientation of the relaxor ferroelectric (FE) $[Pb(Mg_{1/3}Nb_{2/3})O_3]_{(1-x)}$-$[PbTiO_3]_x$ (PMN-PT) with x=0.32 near the morphotropic phase boundary.[13] Using these substrates, changes in magnetization of polycrystalline metallic films have been induced with an applied electric field and can be correlated to the anisotropic in-plane strain generated from FE domain reconfiguration.[14] This phenomenon can be used to re-orient the magnetization of sub-micron nanostructures with strong shape-induce magnetic anisotropy.[15] Furthermore, the use of an 80-100 nm thick single crystalline manganite layer in lieu of the polycrystalline film exhibits a strong strain-based magnetoelectric effect through the rotation of in-plane magnetic anisotropy by up to 22 degrees, as well as hysteretic resistivity tunable though the strain state of the film.[16,17] An analysis of the reversible electric-field switching behavior of the manganite/PMN-PT system through X-ray spectroscopy techniques as well as theoretical calculations showed that the induced strain change leads to a change in Mn $e_g$ orbital population preference, causing a reversible shift in Curie temperature of up to 10 K.[18]

We show here that the model epitaxial perovskite system consisting of a 17 nm thick (011)-oriented $La_{0.7}Sr_{0.3}MnO_3$ (LSMO) film on a PMN-PT substrate can exhibit a robust room temperature magneto-electric coupling manifested as non-volatile changes in magnetic domain structure, magnetic anisotropy, and resistivity due to strong and anisotropic strain-based interactions. To probe such changes, we use chemically and spatially resolved techniques such as X-ray magnetic circular dichroism (XMCD) spectroscopy and X-ray photoemission electron microscopy (PEEM) under the action of constant applied electric fields as well as electric and magnetic field pulses. From these measurements, we directly correlate FE domain re-configuration and thus dynamically tuned film strain state with large and abrupt changes in the magnetic domain structure as well as the resistivity of the LSMO film.

Following the notation of Wu *et al*,[14] we distinguish between the eight possible variants of FE domain orientation in the following manner: the four <111> FE orientations that lie wholly in the (011) plane belong to the $P_{xy}$ poling state, and the four <111> orientations that lie partially out-of-plane are either termed as $P_{z+}$ or $P_{z-}$ poling state depending on the direction of the out-of-plane FE component. An electric field pulse of approximately 2 kV/cm is sufficient to rotate the FE polarization from $P_{xy}$ to $P_z$ or vice versa. Due to the small Thompson-Fermi screening length of order one unit cell compared to the full film thickness,[19] we expect that little change in film behavior will be observed due to charge accumulation or depletion at the film/substrate interface, and thus concentrate on studying the effects of transitions from the $P_{xy}$ to $P_z$ states and vice versa.

An analysis of the change in LSMO unit cell dimensions upon a change in FE domain configuration was performed by poling the substrate into a series of different configurations and measuring the change in position of seven out-of-plane and partially in-plane reflections with high-resolution X-ray diffraction-based reciprocal space mapping. The film peak positions for the $P_{xy}$ configuration were best fit to a monoclinic unit cell, with the average lattice parameter **a** along the [100] substrate direction as 0.3894 ± 0.0002 nm and partially out-of-plane parameters **b** and **c** as both 0.3890 nm within the error of the measurement. The monoclinic angle between **b** and **c** shows a small change from 90.4° ± 0.1° in the $P_{xy}$ state to 90.2° in the $P_{z-}$ state. The change in PMN-PT and LSMO unit cell dimensions along orthogonal in-plane and out-of-plane directions as a function of substrate poling state is compared to a macroscopic strain gauge measurement in Table 1. Differences between the changes in lattice parameter from X-ray diffraction results as compared to the macroscopic strain gauge results from Ref. 13 may be due to partial loss of strain transfer through strain gauge adhesive or similar effects.

| *Crystallographic Direction* | *Strain gauge on PMN-PT* | PMN-PT substrate | LSMO film |
|---|---|---|---|
| In-plane [100] | *-150 ppm* | -400 ppm | -170 ppm |
| In-plane [01$\bar{1}$] | *1300 ppm* | 2300 ppm | 2200 ppm |
| Out-of-plane [011] | | -1100 ppm | -800 ppm |

Table 1 – Spatially averaged change in dimension along orthogonal directions of a PMN-PT substrate measured between the $P_{xy}$ and $P_z$ configurations in zero electric field as measured by a strain gauge (from Wu *et al*.[13]) compared to the change in the same directions determined from X-ray diffraction reciprocal space maps for the LSMO/PMN-PT sample.

While the 17 nm thick LSMO film has a significant static distortion compared to the bulk pseudocubic parameter of 0.387 nm, the film is partially relaxed due to the large mismatch with the PMN-PT substrate (pseudocubic lattice parameter of approximately 0.387 and 0.402 nm, for LSMO and PMN-PT, respectively). However, a large reversible change in LSMO lattice parameter compared to its bulk values is seen when the substrate FE state is changed from $P_{xy}$ to $P_{z-}$, with an anisotropic strain change of $(\varepsilon_{100}, \varepsilon_{01\bar{1}})$ = (0.41%, 0.67%) for the $P_{xy}$ state and (0.43%, 0.44%) for the $P_{z-}$ state. The LSMO film is in tension along both the [100] and [01$\bar{1}$] crystallographic directions, but the change in magnitude of tension is most significant along the [01$\bar{1}$] direction when changing between the $P_{xy}$ and $P_z$ states (see supplemental information). A static anisotropic distortion is seen in films grown on dielectric substrates due to the anisotropic elastic moduli of bulk (La,Sr)MnO$_3$[20], but the large change in dilational strain along the [01$\bar{1}$] direction allows for tuning of the film strain from a highly anisotropic to a nearly isotropic strain state.

In contrast to previous studies of almost fully-relaxed LSMO films on PMN-PT (thickness = 80-100 nm),[16,17] the current work displays a static distortion of the LSMO unit cell in the as-grown state in addition to the imprinted strain generated upon reorientation of the substrate FE domains. To disentangle the effects of epitaxial mismatch from electrostrictive or ferroelastic domain contributions due to the applied electric field, we first examine macroscopically averaged functional properties such as sheet resistivity and magnetization and compare these to previously reported studies. Due to the double exchange mechanism responsible for the ferromagnetic order in manganites, there is a coincident metal-insulator transition (MIT) accompanying the paramagnetic-ferromagnetic transition at the Curie temperature ($T_c$).[21] Both finite size effects as well as epitaxial strain as derived in the Millis model[22] may lower the Curie temperature of LSMO as compared to the bulk value of approximately 370 K. For the 17 nm LSMO film, the metal-insulator transition ($T_{MIT}$) for the as-grown state in zero magnetic field is 322 ± 1.3 K and the $T_c$ as measured by superconducting quantum interference device (SQUID) magnetometry is 322 ± 1.6 K. Furthermore, there is a lowering of $T_c$ by approximately 3 K when the sample is poled from the $P_z$ state to the $P_{xy}$ state. The suppression of $T_c$ and $T_{MIT}$ due to the static epitaxial strain or finite size contributions can be expected due to the large lattice mismatch between the LSMO thin film and PMN-PT. However, we can clearly resolve the influence of substrate FE domain configuration on the ferromagnetic and resistive properties through sequential poling of the substrate at room temperature.

In Figure 1, the sheet resistance of the LSMO film in the metallic state at 298 K is shown as a function of applied electric field. Both a major hysteresis loop (±2.5 kV/cm, showing transitions from $P_{z-}$ to $P_{xy}$ to $P_{z+}$ states) and a minor loop (-2.5 kV/cm to 1.8 kV/cm, showing transitions only between $P_{z-}$ to $P_{xy}$ states) are plotted. Peaks in the major loop correspond to FE axis rotations from $P_z$ to $P_{xy}$ or vice versa, causing an increase in in-plane strain and thus an increase in Mn-O bond distance as exhibited by the increase in $\varepsilon_{01\bar{1}}$ measured from reciprocal space mapping. Due to this change, the Mn $e_g$ hopping integral decreases and the resistivity increases. Comparison between the major and minor loops illustrates that the bistable strain at zero applied electric field can produce significant (3.8% change from $P_{z-}$ to $P_{xy}$ in Fig. 1) non-volatile room temperature changes in resistivity. The magnitude of the peaks in resistivity of up to 4% is an increase over previously reported values of approximately 1%, likely due to the large electric-field induced changes in unit cell dimensions over those seen in thicker films.[17] Thus, in spite of the large static epitaxial strain contribution, the significant change in unit cell dimensions also manifests as a significant change in room-temperature resistivity.

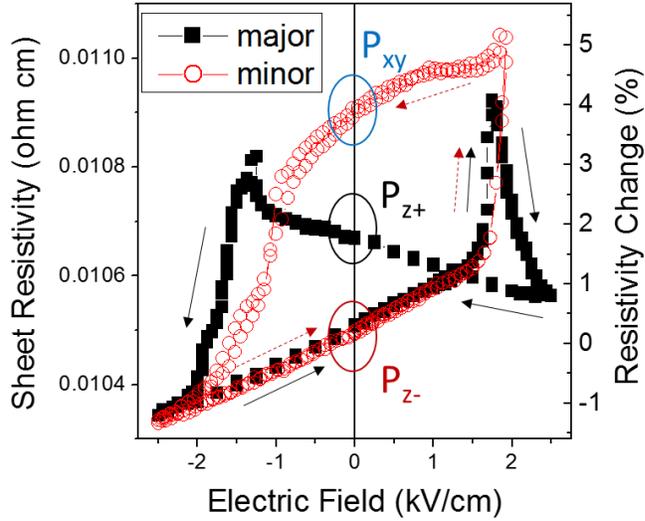

Figure 1 – Electric-field dependence of the LSMO film sheet resistivity at 298 K for a major bipolar loop sweeping from the $P_{z-}$ state to the $P_{z+}$ state as well as a minor loop poling from the $P_{z-}$ to the $P_{xy}$ state over two cycles, showing clear hysteresis in resistivity at zero applied electric field. Solid arrows indicate sweep direction for the major loop, and dashed arrows indicate sweep direction for the minor loop.

The sensitivity of the LSMO resistivity to strain is illustrated by the measurable slope between -2.5 kV/cm and 0 kV/cm where no significant FE axis reorientation takes place but there is a small change in FE axis angle due to the converse piezoelectric effect. The major loop is partially symmetric with respect to electric field polarity, and any asymmetry may be due to the asymmetric electrode configuration (LSMO//PMN-PT//Au) and non-symmetric ferroelectric reversal processes due to the different electrode-ferroelectric interfaces.

Soft X-ray absorption-based magnetic probes offer element sensitivity to magnetization in a non-contact geometry, allowing for the sample to be arbitrarily poled with electric or magnetic field pulses and resulting changes in magnetic properties determined from those external fields. For example, macroscopically averaged XMCD hysteresis loops taken at the Mn $L_3$ edge (Figure 2) show that a large two-fold in-plane magnetic anisotropy exists for the $P_{xy}$ state, with the easy axis corresponding to the [100] in-plane direction and a coercive field of 4 mT. Loop squareness as calculated from the ratio of the magnetization at remanence to the saturation magnetization (S = $M_{rem}/M_{sat}$) is high along the [100] direction, but is significantly reduced along the [01$\bar{1}$] direction as shown in Figure 2. SQUID magnetometry measurements at 300 K verify the trend with XMCD hysteresis loops, with $S_{100} = 0.59$ and $S_{01\bar{1}} = 0.5$ for the out-of-plane poled $P_z$ state 10 weeks after a pulsed field of 6 kV/cm, showing clear retention of the magnetic anisotropy for a significant duration after the electric field pulse.

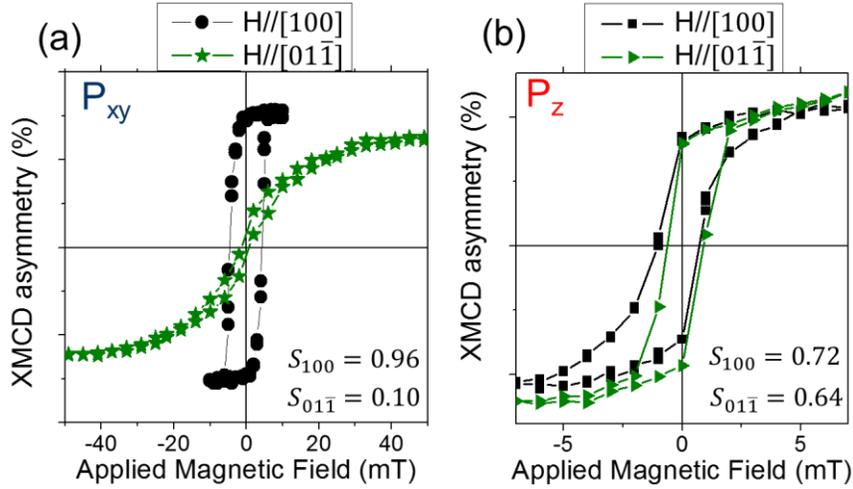

Figure 2 - Mn $L_3$ XMCD hysteresis loops taken at 298 K measured in zero electric field with the substrate poled in the (a) $P_{xy}$ and (b) $P_{z+}$ states. The hysteresis loops are acquired with the magnetic field along two orthogonal in-plane directions, [100] and [01$\bar{1}$]. The [100] direction has high remanence and loop squareness S = $M_{rem}/M_{sat}$ when poled in the $P_{xy}$ state, but the sample is nearly isotropic when poled in either the $P_{z+}$ or $P_{z-}$ state.

This anisotropic behavior is similar to that of (011)-oriented LSMO films grown on SrTiO$_3$ substrates, with a 250 nm thick film having a uniaxial magnetoelastic anisotropy constant of K=8.4 x 10$^4$ erg/cm$^3$.[23]  For the $P_{xy}$ poling state (Fig 2(a)), the hard [01$\bar{1}$] hysteresis loop can be used to estimate an effective anisotropy constant of $K_{eff}$=3.3 x 10$^4$ erg/cm$^3$. However, upon rotation of the FE domains to the $P_z$ state (Fig. 2(b)), the hysteresis behavior is more isotropic and the effective anisotropy constant is reduced to below 8 x 10$^3$ erg/cm$^3$.  As the strain along the [100] direction is not changed significantly by reorientation of the PMN-PT domains as shown in Table 1, we correlate this change in magnetic anisotropy to the reduction in strain along the [01$\bar{1}$] direction for the $P_z$ FE domain state.

The spatially-averaged hysteresis loops allow for a quantitative comparison of the average film anisotropy, while spatially-resolved PEEM measurements allow for local mapping of individual magnetic domains and their response to either electric or magnetic field pulses.  To gain a better understanding of the magnetic domain evolution, we follow a magnetic field protocol similar to that of the ascending branch of a hysteresis loop; saturate the sample at a large negative magnetic field (-37 mT) and observe the domain evolution upon pulsing of small but increasing positive magnetic fields (0.1 to 7.8 mT) as shown in Figure 3.  The sample must be measured in remanence due to the deflection of secondary electrons used for imaging in PEEM by any large external magnetic fields.

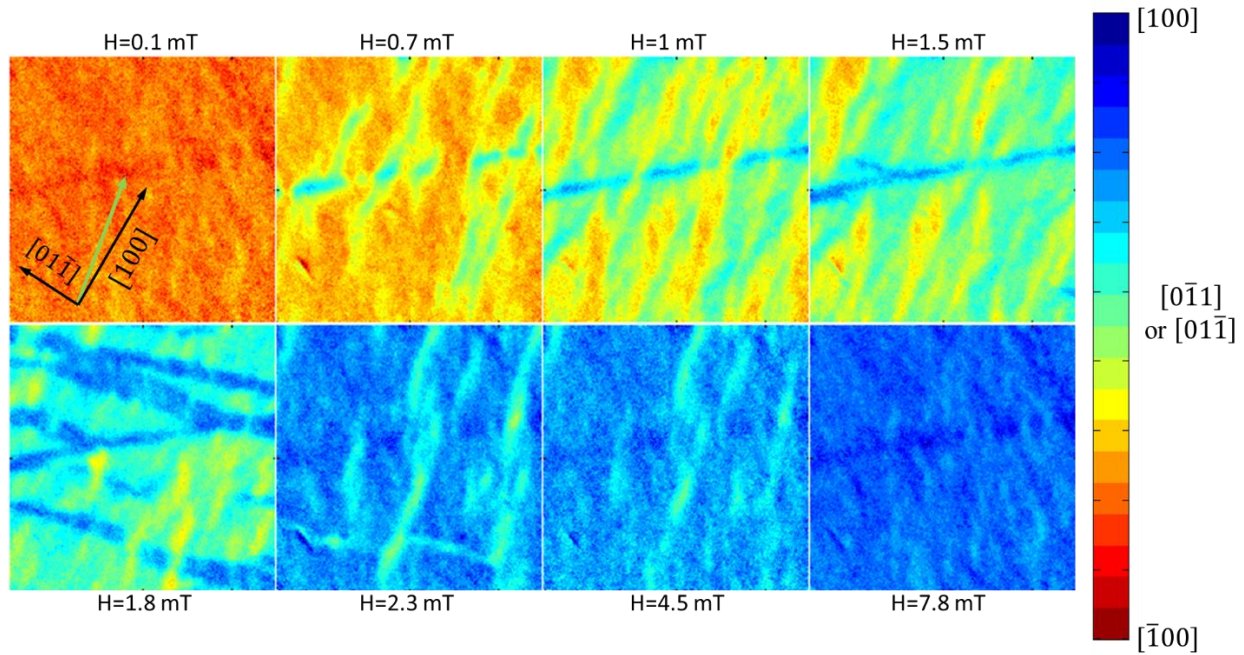

Figure 3 – Montage of colorized PEEM asymmetry images in at 15 x 15 μm region as a function of applied magnetic field pulses with the substrate poled in the $P_{z+}$ configuration and initial applied magnetic field along the $[\bar{1}00]$ direction. The magnetic field pulse and X-ray incidence direction are collinear and are indicated by the green arrow in the upper left panel, with in-plane crystallographic directions shown for reference. Magnetization reversal occurs first by nucleation of many small domains with contrast along the $[01\bar{1}]$ or $[0\bar{1}1]$ directions, followed by domain wall motion and eventual rotation of magnetization towards the [100] direction.

Figure 3 is a series of PEEM images for a $P_{z+}$ poled state with a color scale proportional to the magnetization direction derived from XMCD asymmetry measurements and is used to distinguish between magnetization strongly aligned along the X-ray propagation direction and nearly collinear with the [100] or $[\bar{1}00]$ directions (blue and red, respectively), and magnetization oriented orthogonal to the X-ray propagation direction and aligned along the $[01\bar{1}]$ or $[0\bar{1}1]$ directions (light green). The initial magnetic field pulse orients the LSMO domains in the field of view along the $[\bar{1}00]$ direction, but careful comparison of the contrast variation in the field of view as well as comparison of the field of view for a $P_{xy}$ poled state reveals that the domains cant away from the $[\bar{1}00]$ direction. This canting angle ranges between 10 and 40 degrees as seen by the orange and yellow contrast in the top left panel of Figure 3, whereas magnetization completely along the $[\bar{1}00]$ direction would have a dark red contrast. This canting of magnetization is seen macroscopically in Figure 2 as the $P_{xy}$ hysteresis loop has $S_{100} = 0.96$, but for the $P_z$ poled state, the remanent magnetization is significantly lower than the saturation magnetization. PEEM vector magnetometry mapping (see supplemental information) confirms a significant canting of in-plane magnetization away from the [100] direction for the $P_z$ state.

A small positive magnetic field of 0.7 mT nucleates many small domains whose magnetization is most closely oriented along the $<01\bar{1}>$ directions. These initial nucleated domains are on the scale of 1-2μm, similar in size to the average domain size of PMN-PT ferroelectric domains as measured by piezoelectric force microscopy.[15, 24] Further field pulses cause domain wall motion along the applied field direction, but after a field pulse of 1.5 mT, a majority of the domains are oriented orthogonal to the applied field. Higher field values are required to rotate the magnetization of these domains to align with the [100] direction, and only after a field pulse of 7.8 mT does the field of view become well oriented towards the [100] direction. Note that this field value is similar in magnitude to the field required to bring the $P_z$ poled sample to saturation in Figure 2. Thus, magnetization reversal occurs through sequential non-180° magnetization rotations for the $P_z$ poled configuration.

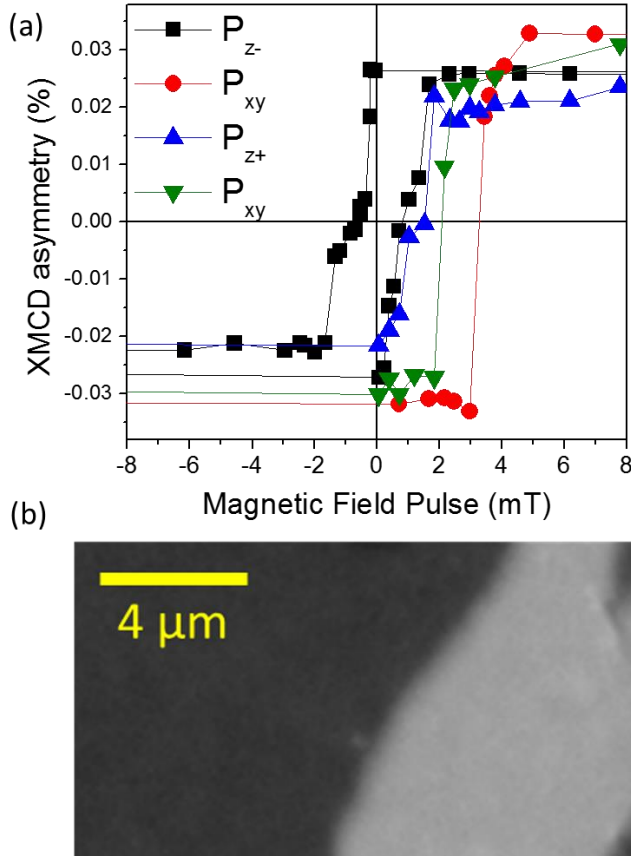

Figure 4 – (a) Median PEEM contrast for a 1 x 1 µm region, equivalent to in-plane magnetization projection 13° from the [100] direction as a function of applied magnetic field pulse and substrate poling state taken from $P_{z-}$ through $P_{xy}$ to $P_{z+}$ and returning to $P_{xy}$. The area is taken from the center of the location shown in Figure 3. (b) PEEM image after a magnetic field pulse of H=3.8 mT showing reversal of magnetization in the field of view by nucleation of a single large domain for the $P_{xy}$ poled state. This region is the same location as the upper half of the field of view shown in Figure 3.

A qualitative comparison may be made between the XMCD hysteresis loops presented in Figure 2 and a spatial average of the PEEM image contrast measured in remanence as plotted in Figure 4. The same 1 x 1 µm region was used for four hysteresis loops, with the sample first poled in the $P_{z-}$ state, poled with a positive field to the $P_{xy}$ state (see supplemental information for corresponding images), poled into the $P_z$ state, then brought back to the $P_{xy}$ state. The top panel of Figure 4 plots the median XMCD contrast for the 1 x 1 µm region, equivalent to the magnetization projection nearly along the [100] direction, as a function of magnetic field pulse. LSMO magnetic domain reversal for the $P_{xy}$ state is markedly different than the series of images shown in Figure 3, with reversal occurring by nucleation of a single 180° reversed area followed by domain wall motion. The stochastically determined field required to nucleate the reversed domain ranges between 3 and 4 mT from the PEEM measurements, which is close to the coercive field obtained from the macroscopically averaged XMCD hysteresis loop in Figure 2. Thus, there is good correspondence between the domain evolution measured in PEEM images taken at remanence with the XMCD hysteresis loops measured in an applied magnetic field. It is the lowered energy barrier for domain nucleation as well as the lowered energy barrier needed to rotate moments to <$01\bar{1}$> directions that leads to the significantly different magnetic hysteresis behavior in the $P_z$ state.

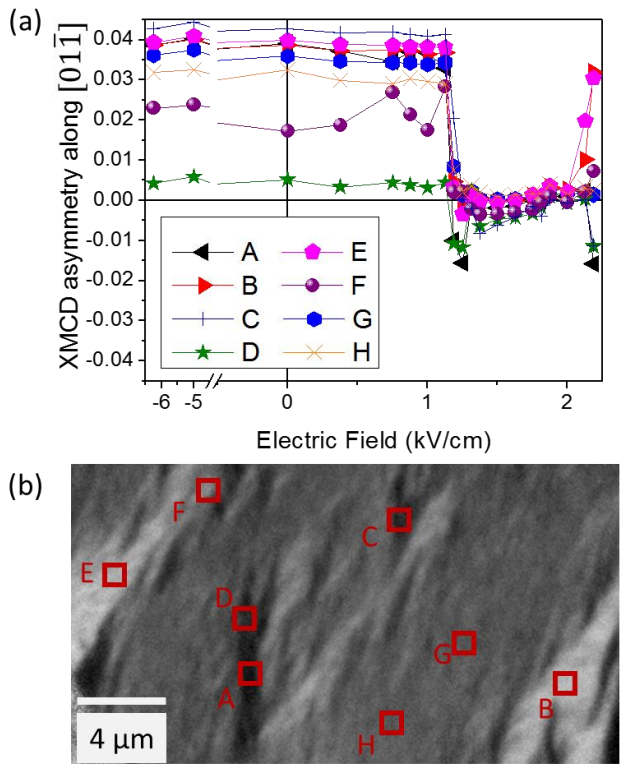

Figure 5 – (a) Comparison of median PEEM asymmetry at 298 K along the [$01\bar{1}$] direction for eight 1 x 1 µm regions (indicated in the bottom panel) on poling the LSMO/PMN-PT sample from the $P_{z-}$ state to the $P_{z+}$ state, showing clear suppression of contrast along the [$01\bar{1}$] direction when the substrate is poled in the $P_{xy}$ configuration. (b) PEEM image taken at a field of 2.2 kV/cm having a mixture of $P_{xy}$ (regions A-F) and $P_{z+}$ (regions G and H) in the field of view.

The bistable nature of the magnetic anisotropy in the LSMO/PMN-PT system offers the possibility of manipulating magnetic domains solely with an electric field. As the anisotropy shifts from nearly isotropic to strongly two-fold symmetry, we postulate that an electric field driven magnetization rotation of 90° or even 180° may be possible. PEEM images with the X-ray incidence direction along the [$01\bar{1}$] are sensitive to rotation of the magnetization from an arbitrary direction towards the <100>

directions as this would result in brighter or darker contrast domains changing to a neutral contrast. By saturating the LSMO sample in a positive magnetic field and sweeping an electric field from -6.5 kV/cm to +2.5 kV/cm, the PMN-PT domain state starts in the $P_{z-}$ state, then switches to $P_{xy}$ and finally is on the verge of the $P_{z+}$ state. Median XMCD contrast for 1 x 1 μm areas are plotted in Figure 5, illustrating a sharp change in contrast at approximately 1.4 kV/cm and 2 kV/cm field values. The initial magnetization state, with a significant number of domains aligned towards the [01$\bar{1}$] direction, is rotated by 90° to lie along the <100> directions at an electric field of 1.5 kV/cm, and upon further poling to the $P_{z+}$ state the magnetization in many domains rotates back towards [01$\bar{1}$] or [0$\bar{1}$1] directions as indicated by both positive and negative contrast above 2 kV/cm. The entire field of view magnetization does not rotate at the same field, but forms stripes of contrast along the [01$\bar{1}$] direction. This result suggests that the PMN-PT FE axis switching field has a large variation, and careful spatially resolved analysis is necessary to distinguish partially switched and fully switched FE configurations. Small changes in strain state between an applied field of -6.5 kV/cm and 1 kV/cm due to the converse piezoelectric effect lead to propagation of domains walls on the scale of 1 μm, suggesting that in addition to the anisotropy tuning obtained from changing strain symmetry, ferromagnetic domain wall manipulation is possible at electric fields smaller than 1 kV/cm.

In order to quantify the angle of magnetization rotation due to the change in film strain, we calculate the contributions to the free energy density[25] stemming from magnetocrystalline (MC) and magnetoelastic (ME) anisotropy terms to determine to what degree a change in strain can induce magnetization rotation in the LSMO film. Magnetostriction constants for manganites in both bulk and thin film form have been found to range from $10^{-5}$ to $10^{-4}$,[23, 26, 27] leading to a significant magnetoelastic anisotropy in the limit of weak magnetocrystalline anisotropy. Figure 6(a) plots the free energy density as a function of in-plane magnetization angle and dilational thin film strain (see supplemental material) and illustrates the significant impact that the anisotropic strain relaxation has on the energy landscape.

In the $P_{xy}$ state (thick red curve), the ME energy term is large compared to the MC term, there are two energy minima at the [100] and [$\bar{1}$00] directions, and a large energy barrier along the [01$\bar{1}$] direction necessitating a large reversal magnetic field for 180° magnetization reversal. On the other hand, the $P_z$ state (blue curve) has only a slight contribution from ME energy and is dominated by the MC energy term (thin black curve). A local minimum at the [01$\bar{1}$] direction suggests that magnetization reversal can occur from [100] to [$\bar{1}$00] through an intermediate 90° rotation with magnetization along a metastable [01$\bar{1}$] direction. Furthermore, the reduction of in-plane strain along the [01$\bar{1}$] direction on changing from a $P_{xy}$ to $P_z$ state (see Table 1) reduces the free energy density for LSMO magnetization along the [01$\bar{1}$] direction by more than 70% (from 1.5 meV/unit cell to 0.28 meV/unit cell). This leads to possible stable magnetization along both <01$\bar{1}$> and <100> type directions for the $P_z$ poled state.

Additionally, shear strains generated due to a change in in-plane unit cell symmetry imposed by the substrate can shift the energy minimum and thus the easy in-plane magnetization direction. For manganite films grown on (011)-oriented $SrTiO_3$ substrates,[28] an in-plane shear strain of up to 0.01 was measured due to templating of the rhombohedral LSMO onto a cubic unit cell. For our LSMO/PMN-PT sample we expect a smaller shear strain due to the non-cubic PMN-PT crystal symmetry as well as the large epitaxial mismatch (see supplemental information), but it is difficult to determine the shear strain from X-ray diffraction measurements due to averaging over an ensemble of many domains. Figure 6(b) shows a magnetization rotation of up to 10 degrees for a modest shear strain of $3 \times 10^{-4}$, smaller than the angular distortion of the PMN-PT surface unit cell generated upon FE axis rotation of approximately $1\times10^{-3}$. Thus, dilational or shear strain changes in the LSMO film can induce a significant change in magnetization angle as well as affect the barrier height for 180° magnetization reversal. Here, we can directly link the change in LSMO unit cell dimensions induced by poling of the PMN-PT substrate from $P_{xy}$ to $P_z$ states or vice versa as listed in Table 1 to the significant changes in magnetic anisotropy symmetry and in-plane magnetic easy axis direction rotation as imaged on the length scale of a single magnetic domain through PEEM.

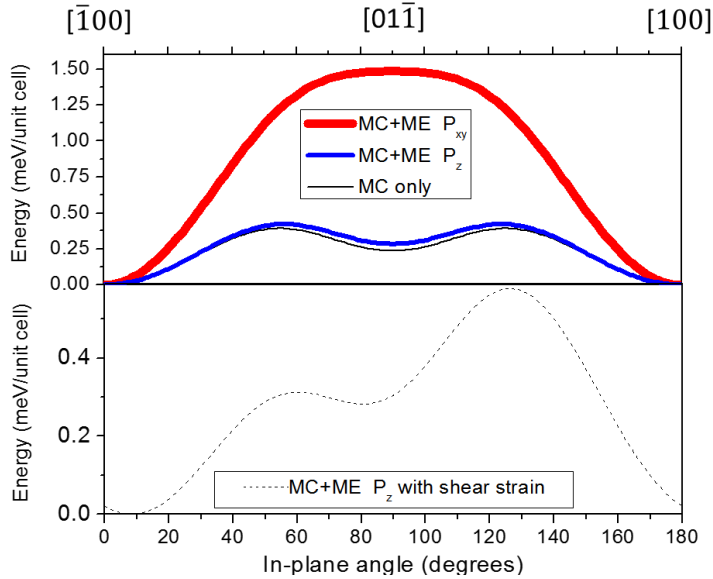

Figure 6 – Change in free energy density as a function of in-plane angle and dilational strain values $(\varepsilon_{100}, \varepsilon_{01\bar{1}})$ determined from X-ray diffraction for the $P_{xy}$ and $P_z$ states. While the $P_{xy}$ state has a uniaxial magnetic easy axis parallel to [100], the $P_z$ state has stable magnetization along both [100] and [01$\bar{1}$] directions, resulting in a weakly four-fold in-plane magnetic easy symmetry. A small shear strain of $3\times10^{-4}$ induces a rotation of the stable magnetization points by 8° and -10° for the [100] and [01$\bar{1}$] stable magnetization directions, respectively.

In conclusion, we have shown that the manganite/titanate artificial multiferroic system can show significant non-volatile room temperature modulation of ferromagnetic domain walls, magnetic anisotropy and resistivity through the careful selection of ferroelectric and ferromagnetic crystal orientation. In particular, domain-resolved imaging of the magnetization shows a marked change in magnetic switching behavior due to the change in imprinted strain state between ferroelectric substrate and ferromagnetic film. Calculation of the change in free energy density due to this imprinted strain state confirms the significant lowering of energy barrier for magnetization reversal, as well as showing that a weak magnetocrystalline anisotropy combined with anisotropic changes in strain are responsible for the change between two-fold and four-fold magnetic easy axes as a function of electric field. This added breadth of functionality when compared to artificial multiferroic systems with polycrystalline metal ferromagnetic elements opens up the possibility of multifunctional low-power room temperature nanostructures and domain-wall devices that take advantage of both of the magnetic anisotropy, domain nucleation barrier lowering, and resistive tuning degrees of freedom and whose functional states may be written and reset with modest electric field pulses.

**Acknowledgements**

Part of this work was performed at the Swiss Light Source, Paul Scherrer Institute, Villigen, Switzerland and the Advanced Light Source, Lawrence Berkeley National Laboratory, USA. This work was partially supported by the EU's 7th Framework Program IFOX (NMP3-LA-2010 246102) and the Defense Advanced Research Projects Agency (Grant N66001-11-1-4135). The Advanced Light Source is supported by the Director, Office of Science, Office of Basic Energy Sciences, of the U.S. Department of Energy under Contract No. DE-AC02-05CH11231.

**Methods**

An LSMO layer of 17 nm thickness was grown on a polished (011)-oriented PMN-PT wafer via pulsed laser deposition. The substrate was held at a temperature of 660°C in a 300 mTorr oxygen ambient, and laser fluence and repetition rate were 1.2 J/cm$^2$ and 1 Hz, respectively. The sample was cooled in a 300 Torr oxygen ambient at 8°C/min. For poling experiments, the LSMO film served as a top contact while a 40 nm gold counter-electrode was sputtered on the back side of the PMN-PT substrate to serve as a bottom contact.

High resolution X-ray diffraction characterization was performed at room temperature using a Bruker D8 Discover system equipped with a monochromatized Cu K$_{\alpha 1}$ source.

Resistivity measurements were performed in a customized Lakeshore TTPX probe station in the van der Pauw geometry with a Kepco bipolar power amplifier as the voltage source for poling the PMN-PT substrate.

The change in free energy from magnetocrystalline and magnetoelastic anisotropy terms as a function of in-plane angle and strain[29-31] were performed by using average values for manganite compliance tensor terms at 300 K ($c_{11}$ =200 GPa, $c_{12}$ = 110 GPa, $c_{44}$ = 45 GPa)[20, 32] and magnetocrystalline anisotropy terms determined from torque magnetometry at room temperature ($K_1$ = 2.6 kJ/m$^3$ and $K_2$ = 5.7 kJ/m$^3$).[23] The in-plane angle locations of stable energy minima were found from the second derivative of the free energy (see supplemental information).

PEEM imaging at the Mn L$_3$ edge was performed at the Surfaces/Interfaces: Microscopy beamline at the Swiss Light Source[33], with magnetic domain structure obtained by taking the XMCD asymmetry of images taken with right and left circularly polarized X-rays $(I_{RCP}-I_{LCP})/(I_{RCP}+I_{LCP})$. X-ray absorption spectroscopy and hysteresis measurements were performed at beamlines 6.3.1 and 4.0.2 of the Advanced Light Source.[34, 35] The LSMO sample surface was kept at 'ground' with respect to standard measurements (- 20 kV for PEEM measurements, -50 V for X-ray absorption measurements), with a custom power supply[36] or Keithley 6487 source applying voltage to the back side gold contact, respectively. The sample was measured in electron yield mode for all measurements, with a grazing incidence angle of 30° from the surface for spectroscopy and hysteresis measurements and 16° for PEEM measurements.

**References**

1. Ramesh, R.; Spaldin, N. A. *Nat Mater* **2007,** 6, (1), 21-29.
2. Heyderman, L. J.; Stamps, R. L. *Journal of Physics: Condensed Matter* **2013,** 25, (36), 363201.
3. Van Den Boomgaard, J.; Van Run, A. M. J. G.; Suchtelen, J. V. *Ferroelectrics* **1976,** 10, (1), 295-298.
4. Ohno, H.; Chiba, D.; Matsukura, F.; Omiya, T.; Abe, E.; Dietl, T.; Ohno, Y.; Ohtani, K. *Nature* **2000,** 408, (6815), 944-946.
5. Roy, K.; Bandyopadhyay, S.; Atulasimha, J. *Applied Physics Letters* **2011,** 99, (6), 063108.
6. Khan, A.; Nikonov, D. E.; Manipatruni, S.; Ghani, T.; Young, I. A. *Applied Physics Letters* **2014,** 104, (26), 262407.
7. Lee, M. K.; Nath, T. K.; Eom, C. B.; Smoak, M. C.; Tsui, F. *Applied Physics Letters* **2000,** 77, (22), 3547-3549.
8. Dale, D.; Fleet, A.; Brock, J. D.; Suzuki, Y. *Applied Physics Letters* **2003,** 82, (21), 3725-3727.
9. Eerenstein, W.; Wiora, M.; Prieto, J. L.; Scott, J. F.; Mathur, N. D. *Nat Mater* **2007,** 6, (5), 348-351.
10. Chopdekar, R. V.; Heidler, J.; Piamonteze, C.; Takamura, Y.; Scholl, A.; Rusponi, S.; Brune, H.; Heyderman, L. J.; Nolting, F. *Eur. Phys. J. B* **2013,** 86, (6), 241.
11. Bowen, M.; Bibes, M.; Barthélémy, A.; Contour, J.-P.; Anane, A.; Lemaître, Y.; Fert, A. *Applied Physics Letters* **2003,** 82, (2), 233-235.
12. Sun, J. Z. *Journal of Magnetism and Magnetic Materials* **1999,** 202, (1), 157-162.
13. Wu, T.; Zhao, P.; Bao, M.; Bur, A.; Hockel, J. L.; Wong, K.; Mohanchandra, K. P.; Lynch, C. S.; Carman, G. P. *Journal of Applied Physics* **2011,** 109, (12), 124101.
14. Wu, T.; Bur, A.; Wong, K.; Zhao, P.; Lynch, C. S.; Amiri, P. K.; Wang, K. L.; Carman, G. P. *Applied Physics Letters* **2011,** 98, (26), 262504.
15. Buzzi, M.; Chopdekar, R. V.; Hockel, J. L.; Bur, A.; Wu, T.; Pilet, N.; Warnicke, P.; Carman, G. P.; Heyderman, L. J.; Nolting, F. *Physical Review Letters* **2013,** 111, (2), 027204.
16. Yang, Y.; Meng Yang, M.; Luo, Z. L.; Huang, H.; Wang, H.; Bao, J.; Hu, C.; Pan, G.; Yao, Y.; Liu, Y.; Li, X. G.; Zhang, S.; Zhao, Y. G.; Gao, C. *Applied Physics Letters* **2012,** 100, (4), 043506.
17. Yang, Y.; Luo, Z. L.; Meng Yang, M.; Huang, H.; Wang, H.; Bao, J.; Pan, G.; Gao, C.; Hao, Q.; Wang, S.; Jokubaitis, M.; Zhang, W.; Xiao, G.; Yao, Y.; Liu, Y.; Li, X. G. *Applied Physics Letters* **2013,** 102, (3), 033501.
18. Heidler, J.; Piamonteze, C.; Chopdekar, R. V.; Uribe-Laverde, M. A.; Alberca, A.; Buzzi, M.; Uldry, A.; Delley, B.; Bernhard, C.; Nolting, F. *Physical Review B* **2015,** 91, (2), 024406.
19. Hong, X.; Posadas, A.; Ahn, C. H. *Applied Physics Letters* **2005,** 86, (14), 142501.
20. Darling, T. W.; Migliori, A.; Moshopoulou, E. G.; Trugman, S. A.; Neumeier, J. J.; Sarrao, J. L.; Bishop, A. R.; Thompson, J. D. *Physical Review B* **1998,** 57, (9), 5093-5097.
21. Goodenough, J. B. *Physical Review* **1955,** 100, (2), 564-573.
22. Millis, A. J.; Shraiman, B. I.; Mueller, R. *Physical Review Letters* **1996,** 77, (1), 175-178.
23. Suzuki, Y.; Hwang, H. Y.; Cheong, S.-W.; van Dover, R. B. *Applied Physics Letters* **1997,** 71, (1), 140-142.
24. Wu, T.; Bao, M.; Bur, A.; Kim, H. K. D.; Mohanchandra, K. P.; Lynch, C. S.; Carman, G. P. *Applied Physics Letters* **2011,** 99, (18), 182903.
25. Hu, J.-M.; Nan, C. W. *Physical Review B* **2009,** 80, (22), 224416.
26. O'Donnell, J.; Rzchowski, M. S.; Eckstein, J. N.; Bozovic, I. *Applied Physics Letters* **1998,** 72, (14), 1775-1777.

27. Srinivasan, G.; Rasmussen, E. T.; Levin, B. J.; Hayes, R. *Physical Review B* **2002,** 65, (13), 134402.
28. Li, Y.; Sun, J. R.; Zhang, J.; Shen, B. G. *Journal of Applied Physics* **2014,** 116, (4), 043916.
29. Gao, Y.; Hu, J.; Shu, L.; Nan, C. W. *Applied Physics Letters* **2014,** 104, (14), 142908.
30. Paes, V. Z. C.; Mosca, D. H. *Journal of Magnetism and Magnetic Materials* **2013,** 330, (0), 81-87.
31. Wang, J. J.; Hu, J.-M.; Chen, L.-Q.; Nan, C.-W. *Applied Physics Letters* **2013,** 103, (14), 142413.
32. Rajendran, V.; Muthu Kumaran, S.; Sivasubramanian, V.; Jayakumar, T.; Raj, B. *physica status solidi (a)* **2003,** 195, (2), 350-358.
33. Le Guyader, L.; Kleibert, A.; Fraile Rodríguez, A.; El Moussaoui, S.; Balan, A.; Buzzi, M.; Raabe, J.; Nolting, F. *Journal of Electron Spectroscopy and Related Phenomena* **2012,** 185, (10), 371-380.
34. Nachimuthu, P.; Underwood, J. H.; Kemp, C. D.; Gullikson, E. M.; Lindle, D. W.; Shuh, D. K.; Perera, R. C. C. *AIP Conference Proceedings: Eighth International Conference on Synchrotron Radiation Instrumentation* **2004,** 705, (1), 454-457.
35. Young, A. T.; Feng, J.; Arenholz, E.; Padmore, H. A.; Henderson, T.; Marks, S.; Hoyer, E.; Schlueter, R.; Kortright, J. B.; Martynov, V.; Steier, C.; Portmann, G. *Nuclear Instruments & Methods in Physics Research Section a-Accelerators Spectrometers Detectors and Associated Equipment* **2001,** 467, 549-552.
36. Buzzi, M.; Vaz, C. A. F.; Raabe, J.; Nolting, F. *Review of Scientific Instruments* **2015,** 86, (8), 083702.

Supplemental Information for "Domain-resolved room-temperature magneto-electric coupling in manganite-titanate heterostructures"

R.V. Chopdekar, M. Buzzi, C.A. Jenkins, E. Arenholz, F. Nolting, and Y. Takamura

A. X-ray diffraction characterization

The poled PMN-PT substrate in this work can be indexed at room temperature to a monoclinic unit cell with a small deviation from orthorhombic (β~89.86° for a similar composition).[1] For a ferroelectric polarization along a [111] direction, there are {011} planes that have a partial projection of the indicated [111] direction and {011} planes that fully contain the [111] direction.  For an applied electric field along the [011] direction, we can examine the possible out-of-plane and in-plane lattice dimensions as a function of poling state and calculate how much change in epitaxial mismatch can be generated by rotation of the ferroelectric (FE) axis from partially out of plane to wholly in the (011) plane.

X-ray diffraction characterization of the LSMO/PMN-PT sample using a lab diffractometer (Bruker D8 Discover) was performed to evaluate the change in film and substrate lattice parameters as a function of applied electric field.  Lattice parameters for the LSMO film determined from reciprocal space maps and electric-field induced changes in unit cell dimensions are presented in the main text. Figure S1 (a) illustrates the change in out of plane {220} peak intensity for the PMN-PT substrate, with the peaks indicated by vertical lines. Here we plot ω-2θ line scans as a function of substrate poling state as the ω-2θ scans clearly show the electric-field induced changes in relative peak intensity and thus changes to the FE domain populations.

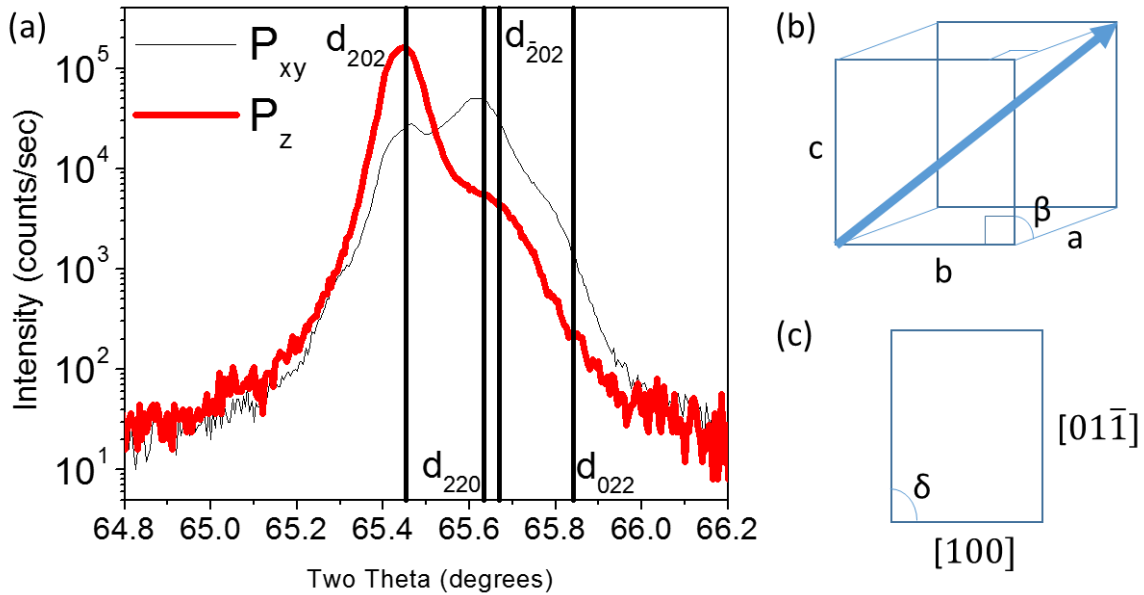

Figure S1 – (a) Out of plane X-ray diffraction ω-2θ scans near the {220} peaks for a PMN-PT substrate as a function of poling state, with vertical lines corresponding to spacings as specified in table S1.  (b) Schematic of a monoclinic cell with a body diagonal indicated by bold arrow and (c) 011 projection of the monoclinic cell.

In Table S1, we tabulate the {220} PMN-PT unit cell parameters found in reference 1, the corresponding 2θ angle, as well as epitaxial mismatch to a fully strained (011)-oriented LSMO. While one might expect all possible orientations present in a thermally randomized sample, we have determined from Figure S1 that the predominant in-plane orientations for the PMN-PT $P_z$ and $P_{xy}$ poling states after electric field cycling correspond to $d_{202}$ and $d_{220}$, respectively ($I_{202}/I_{total}$ = 95% for $P_z$, ($I_{220}$ or $I_{\bar{2}02}$)/ $I_{total}$ = 70% and $I_{022}/I_{total}$ = 5% for $P_{xy}$). The monoclinic orientation in the $P_z$ poling state is straightforward to understand – with a large electric field along the out of plane [011] direction, all FE domains will rotate to align their FE axes with the electric field, and the out of plane [011] length will be large compared to the in-plane $[01\bar{1}]$ direction. For the $P_{xy}$ poling state, the FE axis rotates to lie in the (011) plane, so elongation along either the [100] or $[01\bar{1}]$ directions is possible.

|  | [100] length (Å) | $[01\bar{1}]$ length (Å) | [011] length (Å) | δ (deg) | 2θ (deg) | Epitaxial mismatch along [100] (%) | Epitaxial mismatch along $[01\bar{1}]$ (%) |
|---|---|---|---|---|---|---|---|
| $d_{202}$ | **4.002** | **5.685** | **5.699** | **90** | **65.45** | **-3.09** | **-3.54** |
| $d_{022}$ | 4.034 | 5.669 | 5.669 | 89.9 | 65.84 | -3.86 | -3.27 |
| $d_{\bar{2}02}$ | 4.016 | 5.682 | 5.682 | 89.9 | 65.67 | -3.43 | -3.49 |
| $d_{220}$ | **4.002** | **5.699** | **5.685** | **90** | **65.63** | **-3.09** | **-3.78** |

Table S1 –PMN-PT monoclinic cell data from Ref 1 with the predominant orientations in Figure S1(a) bolded, tabulating in-plane and out-of-plane dimensions as well as in-plane angle, δ. The corresponding diffraction angle from the out-of-plane spacing is also indicated. For comparison, the epitaxial mismatch between (011)-oriented LSMO and PMN-PT unit cells along the orthogonal in-plane directions is also tabulated.

The last two columns of Table S1 show the epitaxial misfit strain between a pseudocubic LSMO unit cell and monoclinic PMN-PT unit cell for each of the possible in-plane {011} planes, and the most significant change between $d_{202}$ and $d_{220}$ is along the $[01\bar{1}]$ direction, whereas a change between $d_{202}$ and $d_{\bar{2}02}$ is along the [100] direction. To first order, we expect an anisotropic strain change in the LSMO unit cell on any single PMN-PT FE domain transitioning from $P_z$ to $P_{xy}$, with a large change in either the in-plane $[01\bar{1}]$ or [100] direction of more than 2400 ppm and little change along the orthogonal in-plane direction. The experimentally derived changes in lattice parameter for the LSMO film presented in Table 1 of the main text suggest that the dominant switching route between $P_z$ and $P_{xy}$ poling states is from $d_{202}$ to $d_{220}$ due to the large change along the $[01\bar{1}]$ direction and negligible change along the [100] direction.

However, this is a simplification of the mismatch between the rhombohedral LSMO unit cell and the PMN-PT in-plane dimensions. For instance, PMN-PT compositions near the morphotropic phase boundary undergo electric-field induced changes in symmetry (e.g. from rhombohedral to orthorhombic), as well as phase transitions due to stress or temperature near ambient conditions.[2,3] Furthermore, the rhombohedral LSMO unit cell forms a microtwin structure when templating on a cubic (110)-oriented SrTiO$_3$ surface, [4] but a change in in-plane shear strain from the as-grown state on PMN-PT can be generated due to the change in in-plane angle δ (see Table S1) between the inequivalent [100] and [01$\bar{1}$] directions. Thus, for the (011) LSMO film, we can expect both elongation type strain along the direction as well as a contribution from shear strain as a function of PMN-PT poling state.

B. PEEM vector magnetometry mapping of the P$_z$ state

A series of PEEM-XMCD images as a function of in-plane azimuthal angle allows for the determination of both the magnitude and direction of the local sample magnetization. Figure S2 shows the in-plane magnetization direction for a sample poled in the P$_z$ state after thermal demagnetization to 340 K and cooling to room temperature in zero magnetic field. PEEM-XMCD images were continuously acquired while heating and cooling to ensure the sample was heated at least 10 K above the point that all magnetic contrast was lost. Magnetic domains align both along the [100] and [01$\bar{1}$] directions to first order, but a more careful comparison of the color levels shows variation in neighboring domains of 10 degrees (e.g. pink vs red domains are mostly oriented along the [100] but are canted away from this direction by ±10°).

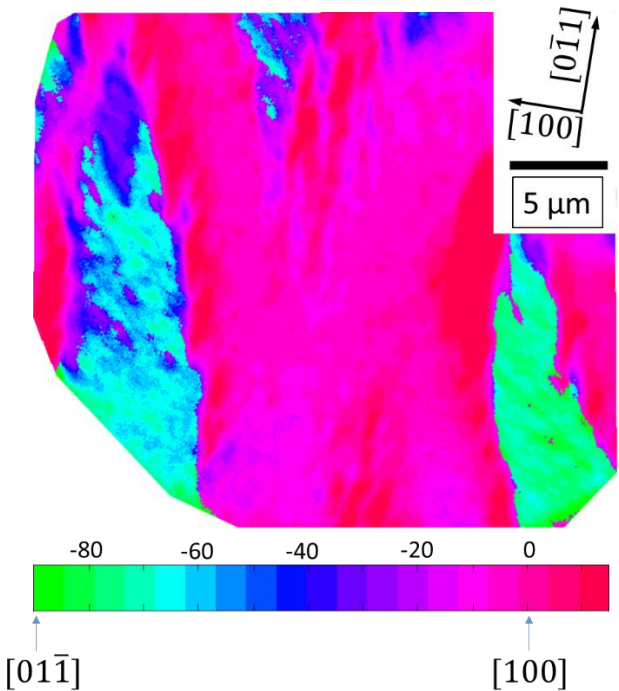

Figure S2 – In-plane magnetization direction for the LSMO thin film with the PMN-PT substrate poled in the P$_z$ configuration. Domains show magnetization both along [100] and [01$\bar{1}$] directions.

C. XMCD images as a function of magnetic field for P$_{xy}$ poling state

In the same sample location as Figure 4, the sample was poled into the P$_{xy}$ state from the P$_{z-}$ state and XMCD asymmetry images were taken during a magnetic field pulse sequence from negative to positive saturation along [100]. The magnetization strongly aligns with the <100> directions in contrast to the $P_z$ state, and the field of view reverses through nucleation of a 180° reversal followed by domain wall motion. A 1 x 1 µm region was integrated as a function of magnetic field pulse and plotted as red circles in Figure 4 of the main text.

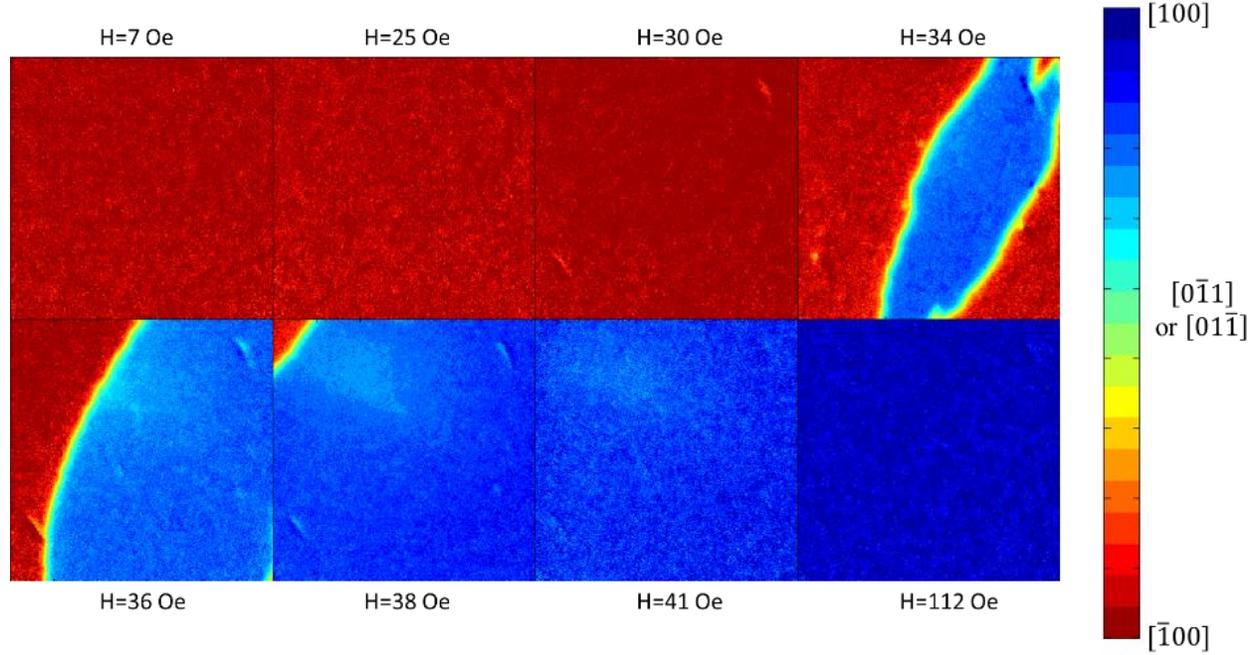

Figure S3 – Colorized XMCD images as a function of magnetic field pulse in the $P_{xy}$ state showing 180° rotation of magnetization, in contrast to the non-180° rotation of magnetization for the $P_z$ state shown in Figure 3.

D. Free energy calculation phase diagram

Figure S4 is a phase diagram of in-plane tensile strain for the (110)-oriented LSMO film with the experimentally determined normal strain values for the $P_{xy}$ and $P_z$ states indicated as points. For this map, we minimize the free energy $f$ from magnetocrystalline and magnetoelastic terms to find stable magnetization angles $\phi$ in the sample plane:

$$f(\epsilon_{100}, \epsilon_{0\bar{1}1}, \phi) = K_1(\alpha_1^2\alpha_2^2 + \alpha_1^2\alpha_3^2 + \alpha_2^2\alpha_3^2) + K_2\alpha_1^2\alpha_2^2\alpha_3^2 + E_{ME}(\epsilon_{100}, \epsilon_{0\bar{1}1}, \lambda_s, \phi)$$

with $\alpha_i$ as direction cosines of the magnetization with respect to the orthogonal in-plane directions of the (011)-oriented film,[5] and the magnetoelastic energy term taken from Gao *et al* for a magnetostrictive film on a (011)-oriented substrate.[6] We first assume negligible in-plane shear strains, compliance tensor components as detailed in the methods section, and an isotropic magnetostriction constant of $\lambda_s = -1 * 10^{-5}$ at 300 K.[7-9]

Examination of anisotropic and bulk magnetostriction constants for LSMO single crystals show that anisotropic effects are negligible within 30 K of the Curie temperature, whereas bulk magnetostriction increases significantly in magnitude near the Curie temperature.[10] The boundaries are generated from the second derivative of the free energy density. The slopes of the phase boundaries are proportional to unity, thus a large anisotropic strain in either direction can induce a strongly uniaxial magnetic easy axis, but nearly isotropic strain allows for the magnetocrystalline anisotropy to dominate and the magnetic easy axis has a fourfold symmetry.

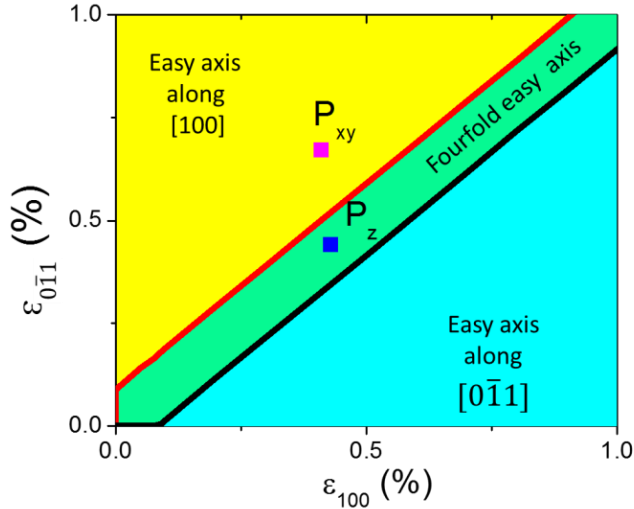

Figure S4 – Magnetic easy axis phase diagram as a function of in-plane strain along the two inequivalent crystallographic directions.

References

1. Singh, A. K.; Pandey, D. *Journal of Physics: Condensed Matter* **2001,** 13, (48), L931.
2. McLaughlin, E. A.; Liu, T.; Lynch, C. S. *Acta Materialia* **2004,** 52, (13), 3849-3857.
3. Peräntie, J.; Hagberg, J.; Uusimäki, A.; Tian, J.; Han, P. *Journal of Applied Physics* **2012,** 112, (3), 034117.
4. Farag, N.; Bobeth, M.; Pompe, W.; Romanov, A. E. *Philosophical Magazine* **2007,** 87, (6), 823-842.
5. Paes, V. Z. C.; Mosca, D. H. *Journal of Magnetism and Magnetic Materials* **2013,** 330, (0), 81-87.
6. Gao, Y.; Hu, J.; Shu, L.; Nan, C. W. *Applied Physics Letters* **2014,** 104, (14), 142908.
7. Darling, T. W.; Migliori, A.; Moshopoulou, E. G.; Trugman, S. A.; Neumeier, J. J.; Sarrao, J. L.; Bishop, A. R.; Thompson, J. D. *Physical Review B* **1998,** 57, (9), 5093-5097.
8. Rajendran, V.; Muthu Kumaran, S.; Sivasubramanian, V.; Jayakumar, T.; Raj, B. *physica status solidi (a)* **2003,** 195, (2), 350-358.
9. Suzuki, Y.; Hwang, H. Y.; Cheong, S.-W.; van Dover, R. B. *Applied Physics Letters* **1997,** 71, (1), 140-142.
10. Demin, R. V.; Koroleva, L. I.; Balbashov, A. M. *Journal of Magnetism and Magnetic Materials* **1998,** 177–181, Part 2, (0), 871-872.